\documentstyle[12pt,aasms4]{article}

\begin{document}

\title{Comparisons of various model fits to the Iron line profile in MCG-6-30-15 }

\author{\bf R. Misra}
\affil{Inter-University Centre for Astronomy and Astrophysics, Pune, India}
\authoremail{rmisra@iucaa.ernet.in}
\author{\bf F. K. Sutaria}
\affil{Inter-University Centre for Astronomy and Astrophysics, Pune, India}
\authoremail{fks@iucaa.ernet.in}

\begin{abstract}
The broad Iron line in MCG-6-30-15 is fitted to the Comptonization model where
line broadening occurs due to Compton down-scattering in a highly
ionized optically thick cloud. These results are compared to the disk line model
where the broadening is due to Gravitational/Doppler effects in the 
vicinity of a black hole.  We find that both models fit the data well and
it is not possible to differentiate between them by fitting only the ASCA data.
The best fit temperature and optical depth of the cloud are found to be $kT = 0.54$ keV and
$\tau = 4.0$ from the Comptonization model. This model further suggests that 
while the temperature can be assumed to be constant, the optical depth varies
during the observation period.  
We emphasis an earlier conclusion that simultaneous broad band data ($3 - 50$ keV) can
rule out (or confirm) the Comptonization model. 

\end{abstract}


\section{Introduction}

A long ($\approx 4.5$ days) observation by ASCA of the Seyfert 1 MCG-6-30-15 revealed
that the Iron line profile in this source is broad with velocity
width of $ > 10^{10}$ cm sec$^{-1}$(Tanaka et al. 1995). A more
detailed analysis of the data by Iwasawa et al. (1996) showed that the profile
is variable and the line width is maximum when the source intensity is minimum. Broad 
Iron lines have also been detected in other AGN by ASCA ( e.g. Nandra et al. 1997).
The most straight forward interpretation of this feature is the disk line model where the line is
broadened due to the combined effects of Doppler and gravitational red-shift in
the vicinity of a black hole (Fabian et al. 1989 ). From modeling the
X-ray data, Iwasawa et al. (1996) showed that 
the disk producing
this line has to have an inner radius $\approx 1.2  r_s$ where $r_s = GM/c^2$ is the
Schwarzschild radius. This would imply that the black hole is spinning close
to its maximal value. If this interpretation is correct then this would be
the first direct observation of the strong gravitational effects expected in
the vicinity of a black hole. This would strongly constrain theoretical models since the inner disk region has to be cold in order to produce the
Iron line while at the same time the hard X-ray producing region would also
have to be located close to the inner edge of the disk. Reynolds \& Begelman (1997) 
have argued that if some of the Iron line
emission arises from  the region inside the last stable orbit, the black hole need not
be a rotating one.

Sulentic et al. (1998) showed there was a disagreement between the inclination
angle derived by fitting the Iron line profile and that from HI and H$\alpha$
measurements. 
Several AGN for which a broad Iron line was detected have the peak centroid
close to $6.4$ keV (Nandra et al. 1997). It was pointed out by Sulentic, Marziani \& Calvani (1998) that this is not expected if the disks are
oriented in random directions. They propose that the Iron line profile is a sum
of two independent Gaussian lines in order to explain the different correlation
relationships of the red and blue side with the source intensity. They also
claimed that there is a blue wing in the Iron line emission which cannot be
explained by the disk line model. These discrepancies and the importance of the
implication of the disk line model warrant a study of alternate models to explain
the phenomenon. Due to the profound implications of
being able to `observe' the immediate vicinity of a supermassive black 
hole, it is important to examine models for the broad Iron line that do
not require strong gravity.

As an alternative to the disk line model, Czerny, Zbyszewska and Raine (1991)
proposed that the line is intrinsically 
narrow and it gets broadened due to Compton down scattering of the photons 
as they
pass through an optically thick cloud. This model is referred to here as the
Comptonization model.
Both, the Comptonization and the disk line models, predict a broad 
Iron line with an extended
red wing.  However, the detailed spectral shape of the line in the 
two models is different.
In particular the disk line model predicts a double peaked line profile
while the Comptonization model predicts a smooth feature. The observed 
line profile in MCG-6-30-15 can be fitted well by 
the disk line model (Iwasawa et al. 1996), 
but so far, no fits of this data have been made to the Comptonization
model. Fitting the data to the Comptonization model would yield values 
for the optical depth and the 
temperature of the cloud -- parameters that can be used to test the physical
feasibility for the existence of such a cloud.

Fabian et al. (1995) rejected the Comptonization model by arguing that the surrounding Comptonizing cloud has to 
have 
a radius $R < 10^{14}$ cms in order that the cloud be highly ionized and
does not produce strong absorption lines. For a $10^7 M_\odot$ black hole
this would imply that the Iron line producing region is smaller than $50 r_s$ and
gravitational effects would be important. The lack of a blue wing in the
best disk-line fit to the line profile, 
implies that the temperature of the cloud should be $kT < 0.2$
keV, which is in apparent conflict with the expected Compton temperature of the cloud.
Further a break in the continuum around $20$ keV is expected for the Comptonization model which
has not been observed. On the other hand,  Misra \& Kembhavi (1998) pointed
out that for a smaller
sized black hole, the intrinsic Iron line produced may not be significantly
broadened by gravitational effects. They calculated the equilibrium temperature
of the cloud to be around $0.2$ keV provided an intense UV source 
is assumed to be
present in this source. They showed that the present broad band data for this source is consistent with the Comptonization model.
 In these arguments the temperature of
the cloud was constrained by the absence of a blue wing in the fitted spectrum
for the disk line model. However, the unfolded spectrum depends on the model
used to fit the data. In other words, the inferred shape of the Iron line profile 
depends on the model used to describe the fit. It is important to obtain
better estimates for the temperature and optical depths of the cloud
in order to verify the arguments presented by Fabian et al. (1995).
In this paper, we directly
fit the observed line profile to the Comptonization model and hence obtain
more realistic values of the cloud parameters.

The Comptonizing model raises several questions regarding the stability,
geometry and dynamics of such a cloud. The origin of such cloud(s) is
unspecified but some arguments for their existence are presented by
Guilbert \& Rees (1988) and Kuncic, Blackman \& Rees (1996). 
Optically thick obscuring clouds may be responsible for the intensity
dip in MCG-6-30-15 (McKernan \& Yaqoob 1998 ; Weaver \& Yaqoob 1998).

The broad line may also be an artifact of simplistic modeling
of the continuum. If the underlying continuum is complex and not
a simple power-law, residuals caused by this fit may have been
interpreted as a broad line. In order to test this hypothesis, we fit
the data  to a narrow line and a broken power-law as the continuum.

In the next section we briefly describe the observations and data analysis
technique, In \S 3 the three models used for the fitting are described. In
\S 4 the results of the data analysis is presented. In \S 5 we discuss the
implications of the results and summarize the paper in \S 6.

\section{Observations}

ASCA observed MCG-6-30-15 from 1994 July 23 to 27 (Tanaka, Inoue \& Holt 1994)
and the data has been analyzed in detail by 
Iwasawa et al. (1996). 
The SIS data which has been analysed in this paper, was obtained in Faint/1CCD mode. 
We used screened events files supplied by the HEASARC online service for this 
observation, where the Faint mode data was already converted into Bright (B) 
and Bright2 (B2) modes for analysis.
In this paper, we analyze
the $3 - 10$ keV data since  below $3$ keV the spectrum is affected by the partially ionized
gas (``the warm absorber'') surrounding the source. Following Iwasawa et al. (1996), 
the data set was divided into three epochs of which the epoch labeled ``high'' encompasses
the highest intensity level of the source. The epoch labeled ``low''
encompasses the lowest intensity
level of the source. The rest of the observation was grouped into a
single data set labeled ``medium''.
Data from both the SIS chips (SIS 1 and 2) for the Bright and Bright2 modes
were grouped and analyzed together.
Only the relative normalization between
these four sets of data  was allowed to vary, but in all cases the variation was found to be less than 2\%.

\section{Brief description of Models}

\subsection{Disk Line model}

The disk line model of Laor (1991) 
assumes that the emission arises from an accretion disk around a black hole 
and includes general relativistic effects.  The inner radius of the accretion disk is $r_i$,
where $r_i = 1.235 r_s$ and $r_i = 6 r_s$ ($r_s = GM/c^2$) correspond
to the case of a maximally rotating black hole and a non-rotating one respectively.
The model is parameterized by $r_i$, the outer radius $r_o$, the emissivity index
$\xi$, the rest frame line energy (fixed at 6.4 keV) and line intensity.  
The emissivity index ($\xi$) specifies the dependence of the emissivity with radius.
The continuum is assumed to be a power-law characterized by two parameters --
photon index $\Gamma$ and a normalization factor. Thus the total number of free parameters used for the
fitting is six.

\subsection{Comptonization model}

The physical geometry of this model is that the central engine consists 
of a hard X-ray source and a cold (UV producing) medium. The X-rays 
impinge upon the cold medium and produce a fluorescence Iron line and
a reflected spectrum. However, unlike the disk line model, the Iron line
is produced sufficiently away from the black hole and is intrinsically narrow
and is broadened by the presence of an extended Comptonizing cloud.
It is important to note
that for self consistency both the intrinsic power-law and Iron line are Comptonized by the
surrounding cloud. In principle, a reflected component which peaks around
$50$ keV  should also be taken into account. 
Since the ASCA observations are restricted to photons
with energy less than $10$ keV we can neglect the reflected component here
(Iwasawa et al. 1996).

The intrinsic spectrum is assumed to be a power-law with photon index
$\Gamma$ and a narrow Gaussian line with $\sigma = 0.05$ keV and centroid energy $E = 6.4$ keV.
The source is assumed to be surrounded by a highly ionized cloud with optical depth $\tau = n \sigma_T R$ at a 
temperature $kT$. The output spectrum is calculated using the Kompane'ets equation. The Comptonization model has seven parameters -- $\tau$, $kT$, the intrinsic photon index $\Gamma$, 
the line energy (fixed at 6.4 keV ), the intrinsic width of the line (fixed at $0.05$ keV), the line
intensity and a normalization factor. Thus the total number of free parameters used for the
fitting is five.
 
\subsection{Broken power-law model}
The broken power-law is an empirical model used to test the hypothesis that the Iron line
in this source is narrow and the artificial broadening observed is due to complexities
in the underlying continuum. The Iron line is assumed to be a Gaussian with width
$\sigma = 0.2$ keV and centroid at $6.4 $ keV. 
The continuum is fit by a broken power-law characterized by a high 
energy photon index $\Gamma_1$, a low energy photon index $\Gamma_2$, a break energy
$E_B$ and normalization.  Thus the total number of free parameters used for the
fitting is five.
  
\section{Results}

The medium intensity (MI) data were fitted to the disk line model of Laor (1991) for 
later comparison with the Comptonization model. The best fit parameters to the
former model are shown in Table 1 and
the unfolded data is plotted in Figure 1.
As mentioned by Iwasawa et al. (1996) this model fits the data well 
with $\chi^2$/ (dof)$ = 1377/(1421)$ (Table 1, first row).
The inner radius $r_i  ( = 2.9 r_s)$ is constrained to be less than $ 3.7 r_s$. However, since the 
line is broader during the low intensity phase, the inner radius is probably close to $1.235$ which corresponds
to a maximally rotating black hole. Setting  $r_i = 1.235 r_s$ and  
refiting the data gave  $\Delta \chi^2 = 1.0$ (Table 1, second row).

The low (LI) and high intensity (HI) data sets were fitted to the
disk line model after setting  $r_i = 1.235$, $r_o = 16.4$ and
inclination angle $\phi = 32^o$. These parameters are not expected to change during the
observation period and freezing them constrains the rest of the  parameters. Moreover the
the number of free parameters become equal to those of the Comptonization model (see below) which 
enables a direct comparison
. The HI data set is formally well fit by the
model with $\chi^2$/ (dof)$ = 1235/(1424)$ (see Table 1). 
There seems to be evidence for an additional narrow component 
(Iwasawa et al. 1996 ; Sulentic, Marziani \& Calvani 1998). 
The emissivity index
($\xi = 2.0^{+0.7}_{-0.7}$) does not vary significantly from the MI state. Much
of the spectral change can be attributed to the slight change in the power-law slope. The line shape during the LI level
is different from that of the MI data (Table 1). The line is broader as indicated by the
increase in the the emissivity index ($\xi = 3.3^{+0.3}_{-0.6}$), since
an higher $\xi$ implies that more photons are produced closer
to the Black hole. The data is again well fit
by the model with $\chi^2$/ (dof)$ = 967/(1424)$. 

\begin{table}
\caption{Spectral parameters for the disk line model of Laor (1991) for the intensity selected
data sets. The parameters are the power-law photon index ($\Gamma$), the emissivity index ($\xi$), the
inner radius ($r_i$) in $r_s$, the outer radius ($r_o$) in $r_s$, the line Intensity ($I_{line}$ in
units of $10^{-4}$ photons/sec/cm$^2$ and the inclination angle $\phi$ in degrees. The line energy is fixed
at $6.4$ keV.}
\begin{tabular}{llllllll}
\hline
Data set & $\Gamma$ & $\xi$ & $r_i$ & $r_o$ & $I_{line}$ &  $\phi$ & $\chi^2$/ (dof) \\
\hline
Medium & $2.05^{+.04}_{-.03}$ & $1.9^{+0.7}_{-0.6}$ & $2.9^{+0.8}_{-1.7}$ & $16.4^{+1.6}_{-1.0}$ & $2.1^{+0.3}_{-0.3}$ & $32^{+1.5}_{-2.0}$ & $1377/(1421)$ \\
Medium & $2.05^{+.034}_{-.03}$ & $1.8^{+0.4}_{-0.2}$ & $1.235$ & $16.4$ & $2.25^{+0.25}_{-0.3}$ & $32$ & $1378/(1424)$ \\
High  & $2.20^{+.06}_{-.06}$ & $2.0^{+0.7}_{-0.7}$ & $1.235$ & $16.4$ & $2.89^{+0.82}_{-0.66}$ & $32$ & $1235/(1424)$ \\
Low  & $2.03^{+.19}_{-.18}$ & $3.26^{+0.31}_{-0.54}$ & $1.235$ & $16.4$ & $2.96^{+0.98}_{-0.91}$ & $32$ & $967/(1424)$ \\
\hline
\end{tabular}
\end{table}

The Comptonization model also fits the MI data well with $\chi^2$/ (dof)$ = 1396/(1423)$ 
although with a higher $\chi^2$ than the disk line model (Table 2). The temperature of
the Comptonizing cloud is found to be $ kT = 0.54^{+0.20}_{-0.19}$ keV. This is hotter than
the temperature estimated by best fit line profile for the disk line model which was 
 $kT < 0.2$ keV (Fabian et al. 1995, Misra \& Kembhavi 1998). The reason for this discrepancy
is that fitting the data to the Comptonization model reveals a blue wing (Figure 2). 
The inferred spectral shape i.e. the unfolded spectra depends upon the model used to
fit the data. Thus the absence of a blue wing in the unfolded spectrum for
the disk line model cannot be used as an estimation of the temperature of the cloud.  
The LI and HI data sets were fitted after setting  $kT = 0.54$. The HI
data is well fit by this model $\chi^2$/ (dof)$ = 1225/(1424)$ (see Table 2). The fit is similar
to the disk line model and again there seems to be evidence for an additional narrow component (Figure 3).
Compared to the MI fit
the optical depth and line strength have increased while the power-law index ($\Gamma$) 
remain unchanged within the limits of error. 
Thus in terms of the Comptonization model, the photon index variation in the
fit to the disk line model can be attributed to the change in degree of Comptonization of the
intrinsic continuum. However,
the limited statistics of the data set and the presence of an additional narrow line does not
allow for any concrete statements to be made. The LI is well fit with a 
larger optical depth $\tau = 6.7$ with $\chi^2$/ (dof)$ = 969/(1424)$. The larger optical
depth leads to a broader line for this set.

\begin{table}
\caption{Spectral parameters for the Comptonization model for the intensity selected
data sets. The parameters are the power-law photon index ($\Gamma$), the optical depth ($\tau$), the
temperature of the cloud ($kT$) in keV and the line Intensity ($I_{line})$ in
units of $10^{-4}$ photons/sec/cm$^2$. The line energy is fixed
at $6.4$ keV and the intrinsic line width is assumed to be $\sigma = 0.05$ keV.
The difference in $\chi^2$ between the disk line (Table 1) and Comptonization model is
tabulated in the last column.}
\begin{tabular}{lllllll}
\hline
Data set & $\Gamma$ & $\tau$ & $kT$ & $I_{line}$  & $\chi^2$/ (dof) & $\Delta \chi^2$\\
\hline
Medium & $1.93^{+.05}_{-.015}$ & $4.16^{+0.33}_{-0.49}$ & $0.54^{+0.2}_{-0.19}$ & $2.22^{+0.55}_{-0.36}$ & $1396/(1423)$ & +18 \\
High  & $1.96^{+.12}_{-.05}$ & $5.32^{+0.62}_{-0.94}$ & $0.54$ & $3.71^{+1.07}_{-1.03}$ & $1225/(1424)$ & -10 \\
Low   & $1.76^{+.18}_{-.44}$ & $6.69^{+2.37}_{-1.20}$ & $0.54$ & $3.81^{+2.84}_{-1.69}$ &  $969/(1424)$ & +2 \\
\hline
\end{tabular}
\end{table}

\begin{table}
\caption{Spectral parameters for the Broken power-law model for the intensity selected
data sets. The parameters are the high energy power-law photon index ($\Gamma_1$), the low
energy power-law photon index ($\Gamma_2$), the break energy ($E_b$) in keV
 and the line Intensity ($I_{line}$ in
units of $10^{-4}$ photons/sec/cm$^2$. The line energy is fixed
at $6.4$ keV and the intrinsic line width is assumed to be $\sigma = 0.2$ keV.
The difference in $\chi^2$ between the disk line (Table 1) and the Broken power-law model is tabulated in the last column. }
\begin{tabular}{lllllll}
\hline
Data set & $\Gamma_1$ & $\Gamma_2$ & $E_b$ & $I_{line}$  & $\chi^2$/ (dof) & $\Delta \chi^2$\\
\hline
Medium & $2.5$ & $1.77^{+0.03}_{-0.03}$ & $5.81^{+0.13}_{-0.15}$ & $0.48^{+0.09}_{-0.08}$ & $1422/(1424)$ & +44\\
High   & $2.5$ & $1.91^{+0.05}_{-0.06}$ & $5.28^{+0.27}_{-0.12}$ & $0.96^{+0.16}_{-0.23}$ & $1208/(1424)$ & -27 \\
Low    & $2.5$ & $1.45^{+0.10}_{-0.15}$ & $5.30^{+0.29}_{-0.26}$ & $0.27^{+0.19}_{-0.17}$ & $972/(1424)$ & +5\\
\hline
\end{tabular}
\end{table}
 
Since the data is from a narrow band (3 - 10 keV) both spectral slopes of the broken 
power-law model cannot be constrained. Therefore, the MI data is fit to the broken 
power-law model with the high energy photon index fixed at 2.0. This gives an unacceptable
fit with $\chi^2$/ (dof)$ = 1506/(1424)$. An Acceptable fit was obtained when the high energy
photon index was set to 2.5 with $\chi^2$/ (dof)$ = 1422/(1424)$. The HI and LI
data sets can also be fitted by this model provided the high energy photon index is 2.5 (Table 3).

\section{Discussion}

The fit to the disk line model to the MI and LI data set is
marginally better than for the Comptonization model. The reduced 
$\chi^2$ for both the models are less than unity. The reason for this
degeneracy is that the spectral shape of the best fit disk line model
does not have a clear double peaked feature (Figure 1). Instead the red wing of
the line gradually extends to $\approx 5$ keV and such a feature
can also be due Compton down-scattering of photons.  
Thus, the narrow band (3 - 10 keV) data used here cannot differentiate 
between these two models. 

 Although the HI data set is formally well fit by the models, there is evidence
for an additional narrow Iron line (Iwasawa et al. 1996) in the residual plots.
This could be due to a real additional narrow Iron line ( Sulentic, Marziani \& Calvani  1998) or because the Iron line shape is variable in
a time scale shorter than the observation time ($\approx 10^5$ secs). 
In the latter scenario the observed shape of the Iron line would be the
time-averaged profile and therefore could not be described by
time-independent modeling. 

Since there are no strong X-ray absorption features in this source the size of the Comptonizing
cloud has to be smaller than $R < 10^{14}$ cms ( Fabian et al. 1995).
For a $10^6 M_\odot$ black hole this corresponds to about $\approx 300 r_s$. The intrinsic
Iron line produced in this case will not be significantly broadened by gravitational red-shift
or Doppler effects.  However if
the black hole mass is larger, these effects would be important and the observed profile
would be complex combination of Comptonization and gravitational effects. In this work for
simplicity we have assumed that the intrinsic line is narrow ($\sigma = 0.05$ keV). The 
temperature of this cloud can be determined by equating the heating and cooling of the gas by
Compton scattering. As described by Misra \& Kembhavi (1998) the temperature of the gas depends
upon the assumed UV flux of the intrinsic source. In Figure 4, the dashed line shows
the assumed intrinsic spectrum for the source. The solid line is the output spectrum after
the radiation passes through the cloud with optical depth $\tau = 4.0$ (the best fit value
for the medium intensity data). The temperature of the cloud was calculated by balancing the input
radiative power to the output power. The UV flux has been assumed here such that this temperature
is equal to the best fit value $kT \approx 0.55$ keV. The UV luminosity is lower than
that estimated by Misra \& Kembhavi (1998) since the best fit value of 
the cloud temperature for the Comptonization model is $0.5$ keV
and not $0.2$ keV as has been assumed earlier. The shape of the UV bump has been
assumed here to have a black-body like shape. However, the equilibrium temperature is not sensitive
to the spectral shape but rather to the total luminosity in the UV band compared to the X-ray band.
The UV flux in MCG-6-30-15 is highly reddened and hence a UV bump is not directly observed 
for this source (Reynolds et al. 1997). Using the minimum value for the extinction,  a lower
limit of the intrinsic flux ($F_{min}$) has been calculated by Reynolds et al (1997) and is 
plotted in 
Figure 4 for comparison. The UV flux required by this model is about a factor ten larger than
this minimum value. The reprocessed IR emission from covering dust in this source also
indicate that the UV flux is much larger than $F_{min}$ (Reynolds et al 1997). 

The temperature of the gas will react to the changing intrinsic spectrum
 in a time scale set by the time for the photons to diffuse out of the system,
\begin{equation}
t_c \approx ({R\over c}) \tau^2  = 8 \times 10^4 \hbox {secs} ({R \over 10^{14} cm}) ({\tau \over 5})^2
\end {equation}
The dynamical time scale is
\begin{equation}
t_d \approx ({R\over v}) \approx = 6 \times 10^4 \hbox {secs} ({R \over 10^{14} cm})^{1/2} ({M \over 10^6 M_\odot})^2
\end{equation}
where $v \approx (2 G M / R)^{1/2} $ is the average bulk velocity of the plasma. Since this two
time-scales are comparable to the observation time scale , the temperature and the optical depth
can vary during the observations. The data is compatible with no change in temperature during
the observation period with the change in the line profile being attributed to changes
in the optical depth of the cloud ( Table 2).  

The continuum spectrum for energies ($ E > 20$ keV) is greatly affected by the presence
of a Comptonizing medium. For $E > 100$ keV the decrease in flux is nearly half of
what is expected from a simple power-law model ( Figure 4). 
The averaged Seyfert 1 spectrum from OSSE has been 
presented by Gondek et al. (1996). They find that at $E = 100$ keV, there is no significant decrease
in the flux from the extrapolated power-law value. However, the statistics is not
good enough to rule out a 50\% decrease in flux. Moreover,
the averaged spectrum may not represent the actual spectrum of
MCG-6-30-15 at a given time.  
Significant differences in the continuum spectrum is also expected for $20 < E < 100$ keV. 
The analysis in this energy
range is complicated by the presence of the reflection component which is significant
only for $15 < E < 60$ keV. Nandra \& Pounds (1994)
analyzed the Ginga data for this source in $ 2 < E < 18 $ keV range and find that a 
reflection component corresponding to a 
solid angle $\Omega_r \approx 2.0 \pi$ is required to explain the data. 
Recently, Lee et al. (1998) analyzed the RXTE (PCA) data for this source
for the energy band (2-20 keV). They find a reflection component corresponding
to a larger solid angle $\Omega_r \approx 3.0 \pi$. 
Preliminary analysis of the BeppoSax observation for this source ( Matt 1998)
requires a $\Omega_r \approx 2.6 \pi$.
These results
indicate that there is an excess flux at $E > 15$ keV which is in apparent contradiction
to the Comptonization model. This excess can be explained within the framework of the
Comptonization model if the intrinsic reflection component has a higher value of $\approx 1.5$ times the fitted $\Omega_r$ 
(Misra \& Kembhavi 1998). For $E \approx 50$ keV, the difference between the Comptonized power-law
and power-law model is larger and can only be removed by invoking an
extremely large unphysical solid angle for the reflected component 
$\Omega_r > 5 \pi$. MCG-6-30-15 has been observed by the 
BeppoSax PDC instrument which is sensitive in the energy range ($13 < E < 200 $ keV ). Matt (1998) show the count spectrum for this source. However, 
the count flux at $E \approx 50$ keV is uncertain by about a factor of two.
Hence it may not be possible to rule out the Comptonization model by this
observation by BeppoSax. 
Similar longer duration broad-band ($3 < E < 100 $ keV ) coverage of this 
source is required to rule out or confirm the Comptonization model. Such an observation is also possible by HEXTE instrument in RXTE.
An additional complication arises
because the source (and hence the Comptonizing cloud) is known to be variable. Hence
simultaneous observation in this range will be required.

Another explanation for these observations could be that the intrinsic line
is narrow but the underlying continuum is complex i.e. not just a simple
power-law. To test this hypothesis we have fit the data to a narrow line
and a broken power-law. We find that the high energy spectral index has to be
$\approx 2.5$ for this model to fit the data. The recent BeppoSax data 
(Matt 1998) seems to contradict
this result since the spectral index obtained was $\approx 2.0$. Moreover, it
is also not clear which physical process will give a break energy $E_b \approx
5$ keV (Table 3).

\section{Summary}

The disk line model and the Comptonization model can fit the Iron profile of this
source for the low and medium intensity data sets. The width and skewness of the
Iron line profile can be reproduced if the source is surrounded by a cloud with
optical thickness $\tau \approx 4$ and temperature $kT \approx 0.6$ keV. This 
temperature is higher than earlier estimates based on fits using the disk line
model. The cloud will have this temperature provided there is a UV bump in the source
which is not visible due to extinction. 

The Iron line profile could also be narrow if the underlying continuum is complex.
Modeling the continuum as a broken power-law allows for a reasonable fit only if
the high energy photon spectral is $\approx 2.5$ which may be in conflict with the recent BeppoSax observations. 

As pointed out earlier by Misra \& Kembhavi (1998), we emphasis the need
for simultaneous long duration broad band ($3 - 50$ keV) observations 
of this source.
Such observations will be able to rule out
(or confirm) the Comptonization model and will strongly constrain any 
theoretical accretion disk models invoked to explain the AGN phenomena.

\acknowledgements

The authors would like to thank S. Raychoudhary and A. Kembhavi for useful discussions. 
 This research has made use of data obtained from the High Energy Astrophysics Science
Archive Research Center (HEASARC), provided by NASA's Goddard Space Flight Center.

\clearpage

\figcaption{The unfolded spectra and best fit disk line model of Laor (1991). The medium
intensity data from both chips (SIS0 and SIS1) for Bright (B) and Bright2 (B2)
modes are shown separately. The dotted line is the power-law while the solid
line is the sum of the power-law and Iron line profile. The parameters of the fit are given
in Table 1. \label{fig1}}

\figcaption{The unfolded spectra and best fit Comptonization model. The medium
intensity data from
both chips for Bright (B) and Bright2 (B2)
modes are shown separately. The dotted line is the Comptonized power-law while the solid
line is the sum of the power-law and Iron line profile. The parameters of the fit are given
in Table 2.\label{fig2}}

\figcaption{The unfolded spectra and best fit Comptonization model. The High
intensity (I3)data from
both chips for Bright (B) and Bright2 (B2)
modes are shown separately. The dotted line is the Comptonized power-law while the solid
line is the sum of the power-law and Iron line profile. The parameters of the fit are given
in Table 2.\label{fig3}}

\figcaption{ The Comptonized multi-wavelength spectrum for MCG-6-30-15. The dotted line
is the assumed intrinsic spectrum of the source while the solid line is the Comptonized one.
The intrinsic component consists of a UV bump, a X-ray power-law and a reflection component. 
The flux
of the UV bump is chosen such that the equilibrium temperature of the surrounding 
cloud is $kT=0.55$ 
keV. Also shown is the estimated UV lower limit for this source ( Reynolds et al. 1997). Maximum
deviation occurs for $E > 50$ keV. \label{fig4}}  

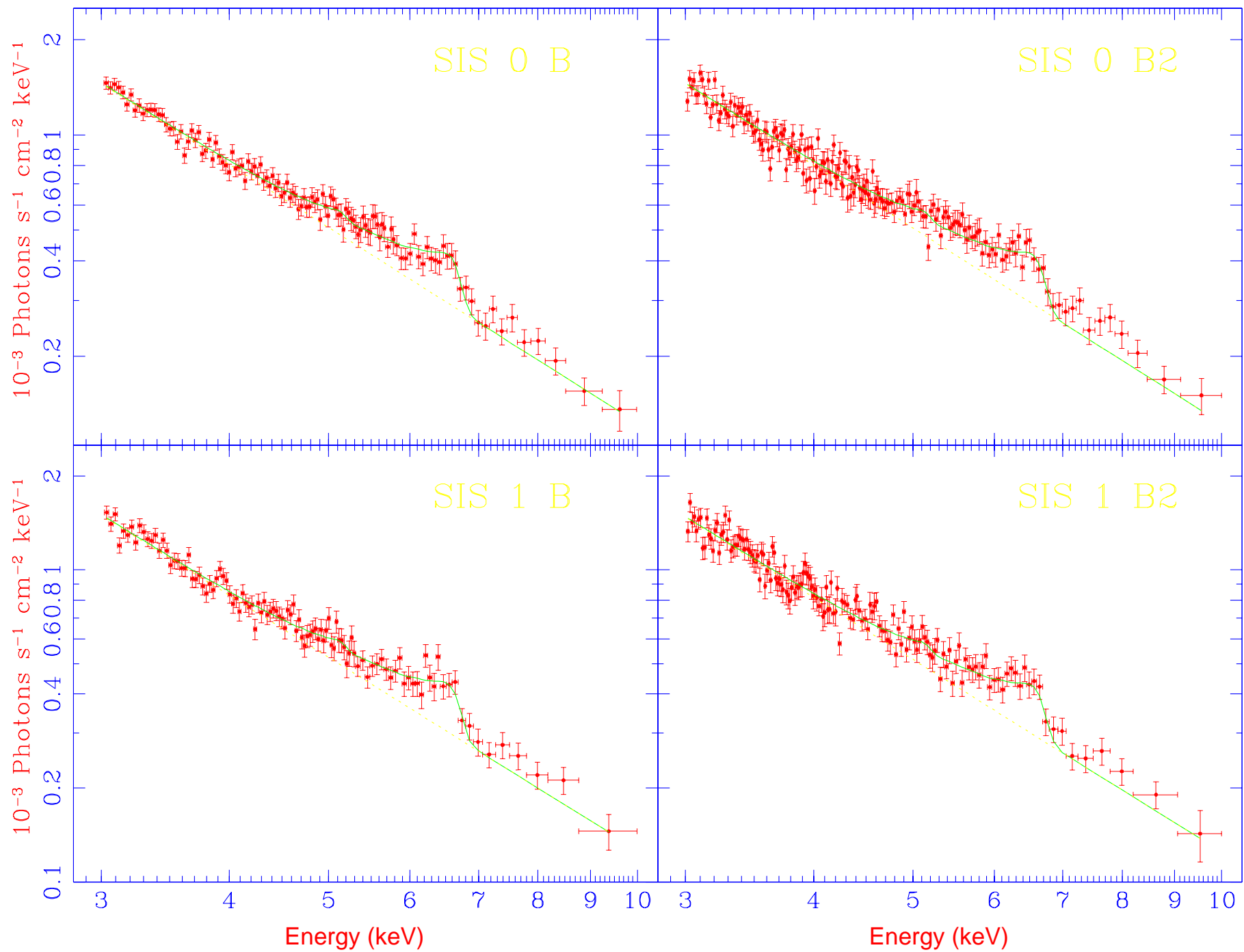
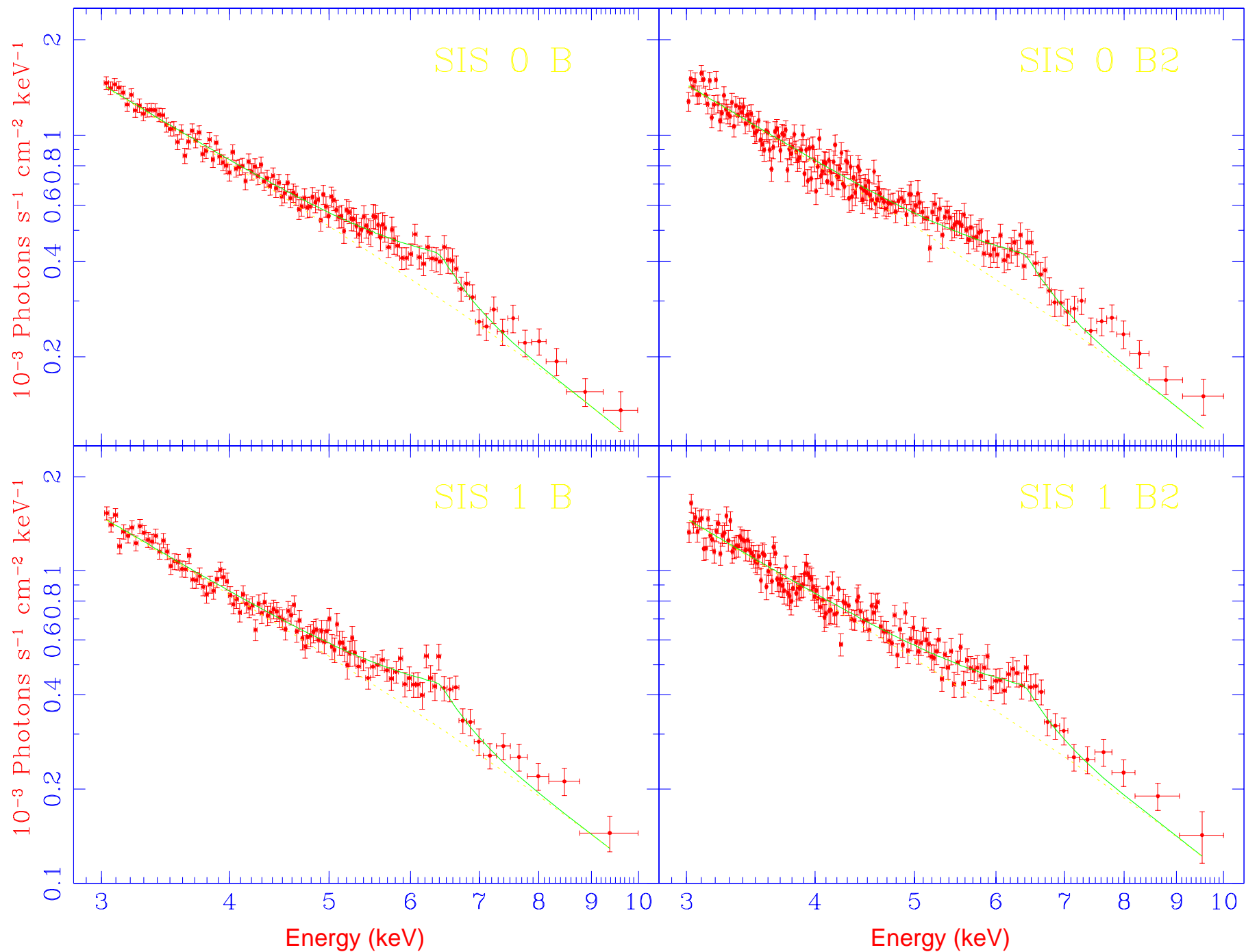
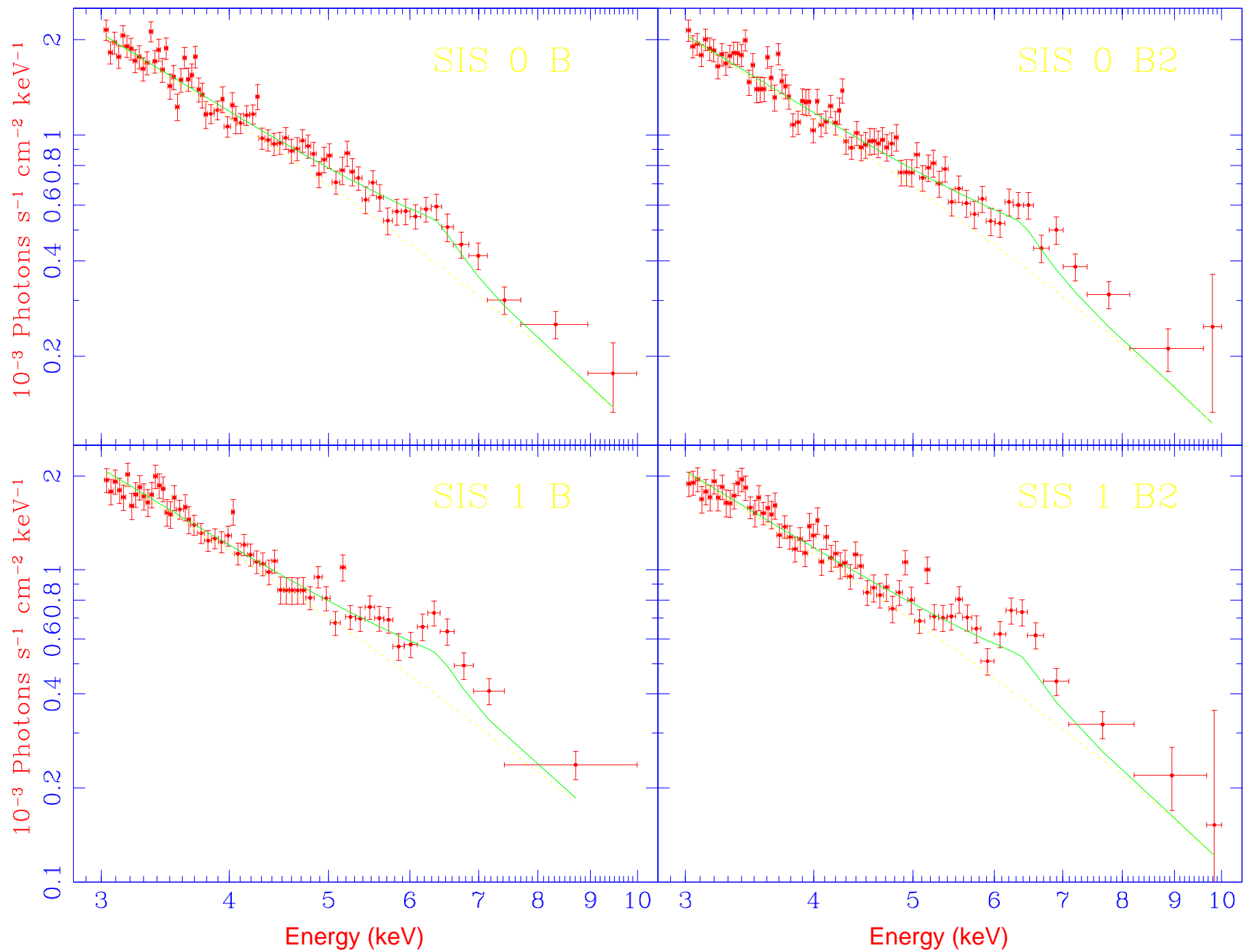
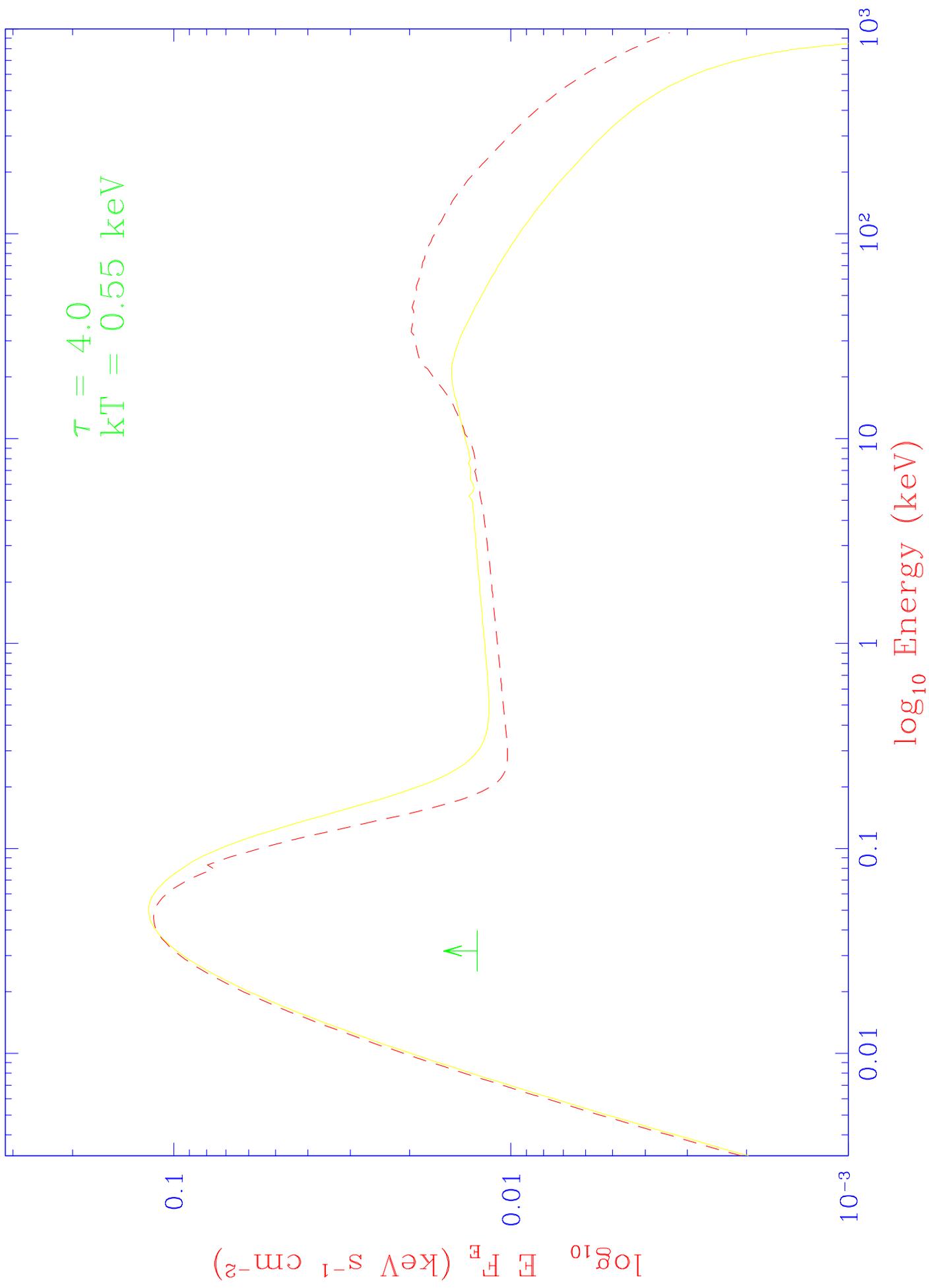
\end{document}